\pdfoutput=1
\documentclass[journal]{IEEEtran}
\usepackage{ifpdf}

\usepackage{cite}

\ifCLASSINFOpdf
   \usepackage[pdftex]{graphicx}
\else
   \usepackage[dvips]{graphicx}
\fi
\usepackage[cmex10]{amsmath}
\usepackage{array}

\ifCLASSOPTIONcompsoc
  \usepackage[caption=false,font=normalsize,labelfont=sf,textfont=sf]{subfig}
\else
  \usepackage[caption=false,font=footnotesize]{subfig}
\fi
 \usepackage{dblfloatfix}
\usepackage{url}

\begin{document}
%
\title{In-Line Post-Process Scribing for Reducing Cell to Module Efficiency Gap in Monolithic Thin Film Photovoltaics}
%
%
%

\author{Sourabh~Dongaonkar,~\IEEEmembership{Student Member,~IEEE,}
        and~Muhammad~A.~Alam,~\IEEEmembership{Fellow,~IEEE}
\thanks{S. Dongaonkar and M. A. Alam are with the School
of Electrical and Computer Engineering, Purdue University, West Lafayette,
IN, 47907 USA e-mail: sourabh@purdue.edu, alam@purdue.edu.}
\thanks{Manuscript received April 19, 2005; revised December 27, 2012.}}

\maketitle

\begin{abstract}
The gap between cell and module efficiency is a major challenge for all photovoltaic (PV) technologies. For monolithic thin film PV modules, a significant fraction of this gap has been attributed to parasitic shunts, and other defects, distributed across the module. In this paper, we show that it is possible to contain or isolate these shunt defects, using the state of the art laser scribing processes, after the fabrication of the series connected module is finished. We discuss three possible alternatives, and quantify the performance gains for each technique. We demonstrate that using these techniques, it is possible to recover up to 50\% of the power lost to parasitic shunts, which results in 1-2\% (absolute) increase in module efficiencies for typical thin film PV technologies.
\end{abstract}


%
\IEEEpeerreviewmaketitle

\section{Introduction}
%
%
%
%
\IEEEPARstart{T}{he} gap between cell and module efficiency is a universal feature of all PV technologies \cite{Green2012a}. This so called ``solar gap'' constitutes about 3-5\% absolute reduction in module efficiency, compared to the corresponding record cell efficiency, for all technologies \cite{Levitan2012}. Many thin film PV (TFPV) technologies have relatively low cell efficiencies; therefore, a significant drop in efficiency at the module level, is often a critical factor limiting their commercial success \cite{}. Moreover, due to the monolithically integrated module manufacturing processes used for TFPV technologies, the problem of cell to module efficiency gap requires a different analysis and design perspective compared to conventional crystalline silicon PV \cite{Hegedus2003}.

Several factors are responsible for the drop in power conversion efficiency from the cell to the module level. A universal loss mechanism arises from the finite sheet resistance of the transparent conducting oxide (TCO) and/or metal layers in TFPV, and the resistances of the emitter and the metal grid lines in crystalline cells. In order to minimize these losses in crystalline PV, a variety of cell and metal grid designs have been proposed \cite{Kerschaver2006,Flat1979}. In monolithic TFPV modules, the resistive losses are minimized by carefully choosing the number of series connected cells, and width of each cell, for each technology type \cite{Johansson2008}. 

Other factors arise from the challenges of depositing thin films over large area ($ \sim m^2 $) substrates in TFPV module manufacturing. These include film thickness variation over large area substrates \cite{Yang2009b}, variation of contact sheet resistance \cite{Hudaya2012}, composition variations in chalcogenide cells \cite{Powalla2003}. These manufacturing challenges not only reduce the module efficiency, but also lead to performance variation at the module level, causing reduced process yield. These may be considered `macroscopic' variations, because they cause performance difference over length scales of several cm, across the module surface. The development of strategies to control these macroscopic variations is the focus of significant process development and research in the TFPV community.

Besides these macroscopic sources of performance variation, there are additional efficiency losses caused by parasitic shunts \cite{dongaonkar:124509} and weak diodes \cite{Koishiyev2009e}. These `microscopic' defects are formed randomly across the solar cell surface, and lead to efficiency loss, when cells are connected in series to form modules \cite{dongaonkar:124509,Koishiyev2009e}. Moreover, these shunts exhibit a log-normal distribution from cell to cell, and cause significant efficiency variation at the module level, which limits the process yield \cite{Dongaonkar2013b}. In crystalline solar cells, it is possible to avoid the random cell efficiency fluctuations caused parasitic shunts, by sorting individual cells into efficiency bins before assembling the modules \cite{Wolden2011}. The monolithic manufacturing process in TFPV, however, means that it is not possible to selectively remove the shunted low efficiency cells in the module. Moreover, the wide cell geometry in TFPV module also leads to significant non-local effects of the shunts on neighboring regions. Thus, a technique of isolating these shunts, after the fabrication of a monolithically integrated module is needed in order to close the cell to module efficiency gap, and to improve process yield.

In this paper, we first describe the 2D SPICE circuit simulation setup in section II, which is used to analyze module performance in presence of parasitic shunts. Then, in section III, we illustrate the effect of random shunts on module performance and variability. Next, in section IV, we discuss the scribing isolation technique for reducing the shunt losses at module level, and use Monte-Carlo simulations to assess their effectiveness. Finally, we conclude, by discussing the practical aspects of implementation of this method.

%
%

\begin{figure}[t]
\centering
\includegraphics[width=3.49in]{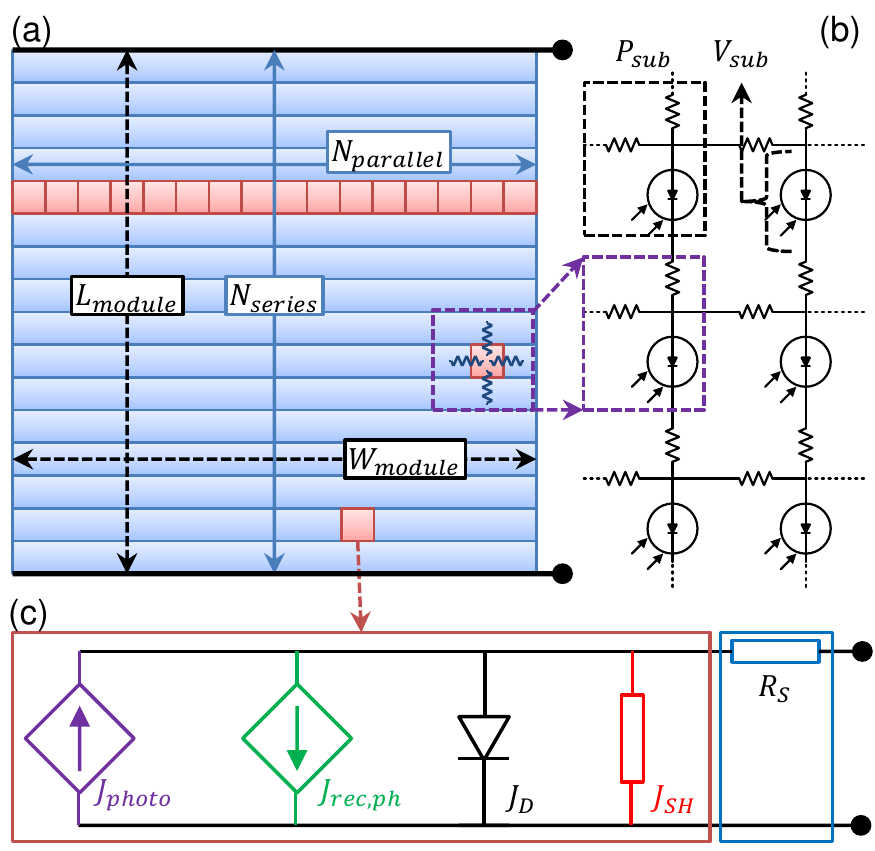}
\caption{(a) Schematic of a typical TFPV module showing module dimensions with $ N_{series} $ series connected cells (rectangles), subdivided into $ N_{parallel} $ sub-cells (squares). (b) These sub-cells are connected using contact sheet resistances to create a 2D mesh representation of the module. (c) Each sub-cell is represented by a physical equivalent circuit for TFPV cells, including the voltage dependent photocurrent, and non-Ohmic shunt current values.}
\label{fig1}
\end{figure}

\section{Simulation Setup}

In order to evaluate the module performance in presence of parasitic shunts distributed across the surface, we use a 2D circuit representation of a TFPV module \cite{Dongaonkar2012} , see Fig. \ref{fig1}(b). This distributed circuit simulation approach is necessary because, the elongated rectangle shaped cells in monolithic TFPV modules leads to important (spatial) interplay of parasitic shunts \cite{Pieters2011}.

Figure 1(a) shows the schematic of a typical TFPV module, with $ N_{series} $ series connected cells. In order to create the circuit representation, we sub-divide each rectangular cell into $ N_{parallel} $ sub-cells. These sub-cells are represented by a physics based equivalent circuit for TFPV cells (Fig. \ref{fig1}(c)), and are connected in series and parallel using the contact sheet resistance values of the TCO and metal layers ($ R_S $). This equivalent circuit accounts for voltage dependent collection ($ J_{photo} $) \cite{Sokel1982}, generation dependent recombination ($ J_{rec,ph} $) \cite{Merten1998}, and the non-ohmic shunt current ($ J_{SH} $) \cite{Dongaonkar2010} components. Each sub-cell, otherwise identical, is assigned different shunt current value, drawn from the log-normal distribution \cite{Dongaonkar2013b}. The details of the simulation setup and validation are described in detail in \cite{Dongaonkar2012,Dongaonkar}. See section S1 in supplementary materials for details regarding equivalent circuit model, and model parameters.

In the following simulations, we use single junction amorphous silicon (a-Si:H) technology as an illustrative example, for ensuring quantitative comparison of the scribing techniques. We use the record cell performance for best sub-cells in the module \cite{Benagli2009}, and the module structure is obtained from manufacturer data sheet \cite{Solar2012}. This simulation setup assumes that the macroscopic sources of variability have been minimized through good process control. Therefore, we can analyze the effect of shunt variability, in particular, using realistic shunt statistics parameters from the literature \cite{Dongaonkar2013b}.

\section{Shunt Statistics and Module Efficiency}

In this section, we discuss the effect of parasitic shunts on the module efficiency. We first analyze the effect of shunted sub-cells on their neighboring regions, by analyzing the distribution of sub-cell voltages and power outputs, obtained from 2D circuit simulations. We then simulate multiple industrial a-Si:H modules ($ L_{module}=104cm $, $ W_{module}=120cm $, and $ N_{series}=104 $) \cite{Solar2012}, in presence of the log-normal shunt distribution, in a Monte-Carlo simulation approach, to quantify the impact of shunt statistics on module efficiency and variability.

\subsection{Non-local Shunt Effect: Illustrative Example}

In order to analyze the effect of shunts in interconnected modules, we simulate a $ 11\times11cm^2 $ sub-module with 11 series connected cells. We first simulate the sub-module with all identical sub-cells, with no parasitic shunts. The resulting IV and PV curves are shown in Fig. 2(b). Note that in this case, the power output of the sub-module is the algebraic sum of the 121 sub-cell power output (including the sheet resistance losses), so that $ P_{submod}^{ideal}=\sum_{121} P_{subcell}^{ideal} = 1.0761W $. 

\begin{figure}[t]
\centering
\includegraphics[width=3.49in]{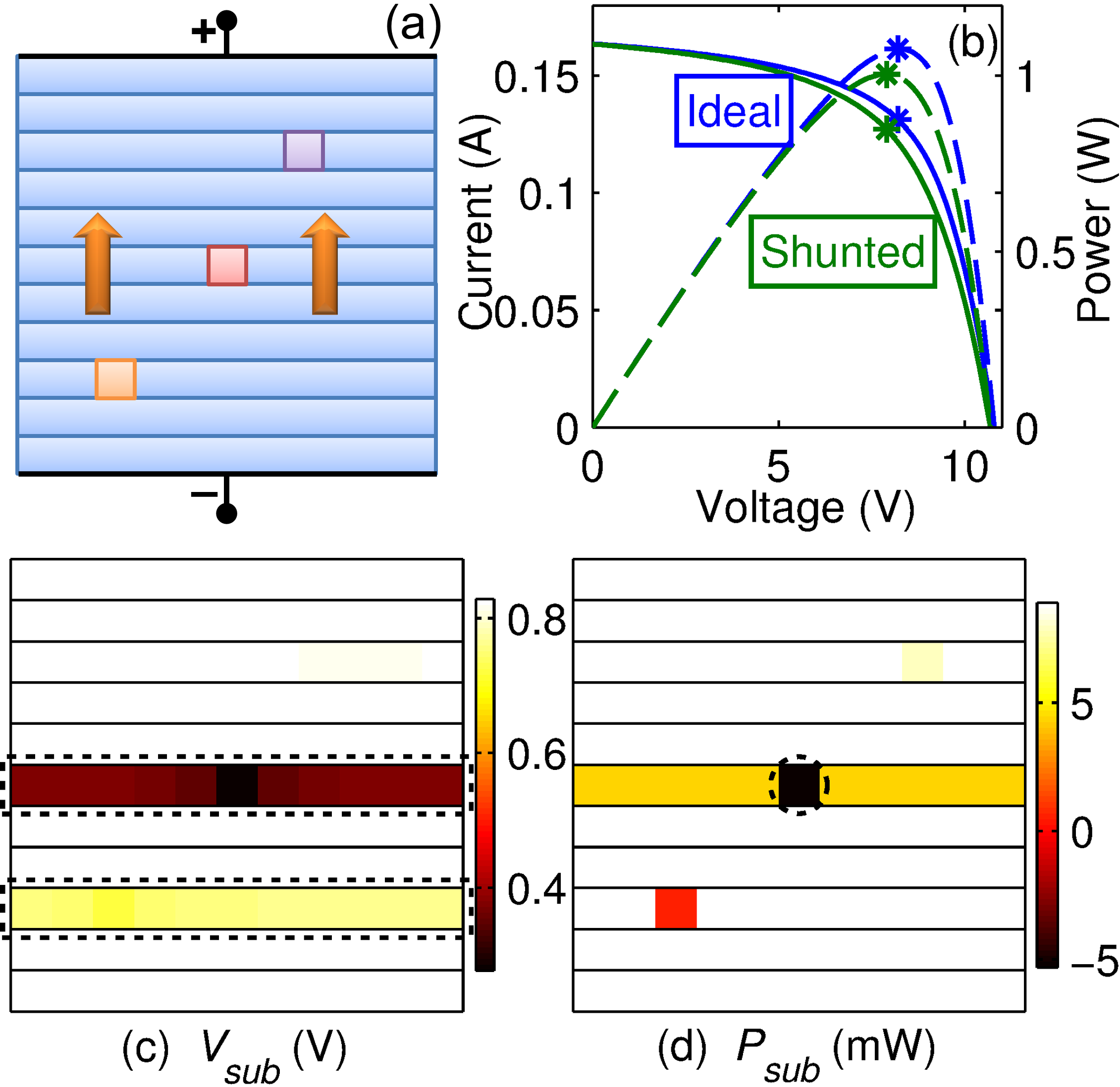}
\caption{(a) Simulation of an $ 11\times11cm^2 $ sub-module, with 11 series connected cells and 3 shunted sub-cells (colored squares), while all other sub-cells are assumed identical. (b) IV (solid) and PV (dashed) curves of the sub-module, without (blue), and with (green) 3 shunts. The symbols denote the MPP values in each case. (c) Distribution of sub-cell voltages $ V_{sub} $ (colorbar in $ V $) at the sub-module MPP (green symbols in (b)), showing that larger shunts (number 3 and 2), lower the operating voltage of their neighboring sub-cells in parallel (highlighted) to $ \sim0.3V $ and $ \sim0.7V $, respectively. (d) The corresponding power output $ P_{sub} $ (colorbar in $ mW $), shows that the sub-cell with largest shunt (number 3), actually consumes ($ P_{sub}\approx-5mW<0 $) the power output of its neighboring regions (circled).}
\label{fig2}
\end{figure}

To demonstrate the implications of shunts on ideal module performance, we now assign 3 different $ I_{SH0} $ values to 3 specific sub-cells, as highlighted in the color plot in Fig. 2(a). The resulting IV and PV curves of the sub-module with these shunts are plotted in Fig. 2(b). As expected, the power output of the sub-module drops in presence of shunts. The magnitude of efficiency loss ($ \Delta\eta\approx-7\% $), however, cannot be explained by simple sum of individual sub-cell outputs. Specifically, we note that even if we assume that the 3 shunted sub-cells produce no power at all, the sub-module output is less than the sum of the power output of the 118 `good' sub-cells; i.e., $ P_{module}^{shunted} (=1.0041W) < \sum_{118}P_{sub}^{ideal} (=1.0494W) $. This means that the shunted sub-cells are modifying the operation of other good sub-cells due to their interconnection, and we must analyze this interaction carefully in order to understand the impact of shunt defects at the module level. 

In order to evaluate this effect of shunts on other sub-cells, we analyze the distribution of sub-cell operating points at the sub-module maximum power point (MPP), marked in Fig. 2(b). Fig. 2(c) shows the operating points of all sub-cells plotted as color plot (color bar is in volts). Note that depending on their magnitude the operating voltage $ V_{sub} $ of the shunted sub-cells, as well as their neighbors is lowered. This is means that sub-cells in parallel to a badly shunted sub-cell, operate away from their MPP voltage, reducing their power output. The number of neighboring sub-cells which suffer from lower operating voltage due to a shunted sub-cell depends on the magnitude of the shunt, as well as the net sheet resistance of the metal/TCO contact layers (compare the highlighted cells in Fig. 2(c)). Note that this effect of shunt defects lowering the operating voltage of the entire cell has been observed experimentally using electroluminescence imaging \cite{Helbig2010,Runai2011}.

Another effect of the interconnection of the shunted and non-shunted sub-cells is that the shunted sub-cell may dissipate power instead of producing it, as seen in the middle cell in Fig. 2(d). In this scenario, the power output of the sub-cells, in parallel to the shunted sub-cell, is dissipated in the defective region instead of flowing to the terminals. Thus, we have a disproportionately large impact on module efficiency due to the effect of shunted sub-cells on their neighbors. First, the sub-cells adjacent to a shunted sub-cell can get biased away from their MPP and produce less power. Moreover, even this reduced power output may be dissipated in the shunted sub-cell. This is because, a heavily shunted sub-cell can get biased so that $ I_{sub} \times V_{sub} < 0 $, and therefore sinks the output current of its neighbors. As a result, even a few shunts can result in a significant reduction in the sub-module output power.

\begin{figure}[t]
\centering
\includegraphics[width=3.4in]{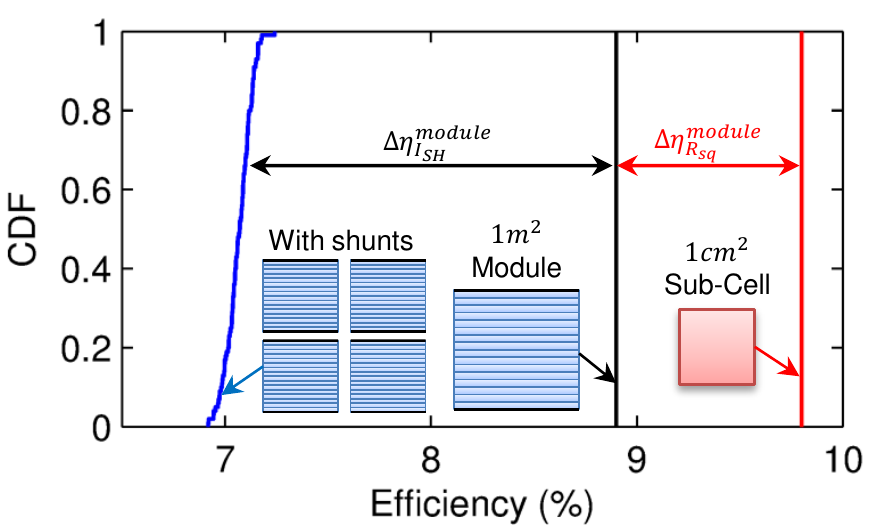}
\caption{CDF plot showing the results of Monte-Carlo circuit simulation of module efficiency (blue), for the a-Si:H technology; in presence of log-normally shunt distribution, across sub-cells. This distribution is compared to the maximum possible module efficiency (black), without any shunts, and the maximum sub-cell efficiency. The efficiency loss due to sheet resistance (red), and parasitic shunt distribution are also highlighted.}
\label{fig3}
\end{figure}

\subsection{Module Efficiency Loss in Realistic Case}

The illustrative example discussed in the previous section explains how relatively few shunts can cause significant reduction in efficiency at a module level. In order to quantify these effects in realistic situations, we need to include the full statistics of shunt formation in thin film solar cells, and simulate full size commercial TFPV modules. It has been shown that shunt current distribution in TFPV cells is described by a log-normal distribution \cite{Dongaonkar2013b}. Each sub-cell is in the module is assigned different shunt current value, from the log-normal distribution. The results of the Monte-Carlo simulations are shown in Fig. \ref{fig3}, comparing the cell efficiency (red line), with the maximum module efficiency without shunts (black line), and the distribution of module efficiencies in presence of shunt variability (blue curve). This shows that if the sheet resistance loss and losses due to parasitic distribution of shunts is accounted for, it can reproduce the typical gap between cell and module efficiency (~3\% absolute here). Note that shunt variability accounts for more than half of this gap (~2\% absolute), and the rest (~1\% absolute) is caused by sheet resistance loss. It is therefore apparent that reducing the effect of parasitic shunts at the module level can significantly improve the module efficiency.


\section{Techniques for Shunt Loss Mitigation}

\subsection{Process Solutions for Shunt Removal}

As parasitic shunt formation is a known issue in all PV technologies, there have been many attempts towards alleviating shunt losses at cell and module level. Some of these methods have focused on process solutions for improving deposition uniformity \cite{Yang2009b}, better substrate cleaning \cite{Johnston2009}, or inclusion of resistive interlayers \cite{Ottoson2006}. Few post-process methods for shunt removal have also been reported. These include, shunt busting in a-Si:H cells \cite{Nostrand1979}, which uses application of high reverse bias on finished cells to burn-up or `bust' the shunt paths. Another approach involves electrolytic treatment of finished solar cell under reverse bias \cite{Nath1988}, or under illumination \cite{Roussillon2004}, for passivating the shunt forming regions, or other non-uniformities on solar cell surface. 

All these techniques focus on modifying the properties of shunt paths, and reducing the leakage through those regions. Moreover, these techniques are indiscriminate, as the whole module is subjected to a particular treatment for shunt passivation, which increases the chances of adversely affecting some other cell property. Finally, these techniques rely on specific, materials, or structures, used in each TFPV technology to achieve shunt passivation, thereby limiting their application more generally. Instead, here we describe a general post-process, scribing isolation technique for mitigating shunt effects, which differs from these earlier approaches in two specific aspects. First, it relies on not removing or passivating the shunts themselves, but on reducing the effect of shunted regions on their neighboring regions. Second, this approach is applicable equally to all TFPV technologies, which use scribing and monolithic integration for module fabrication.

\begin{figure}[t]
\centering
\includegraphics[width=3.49in]{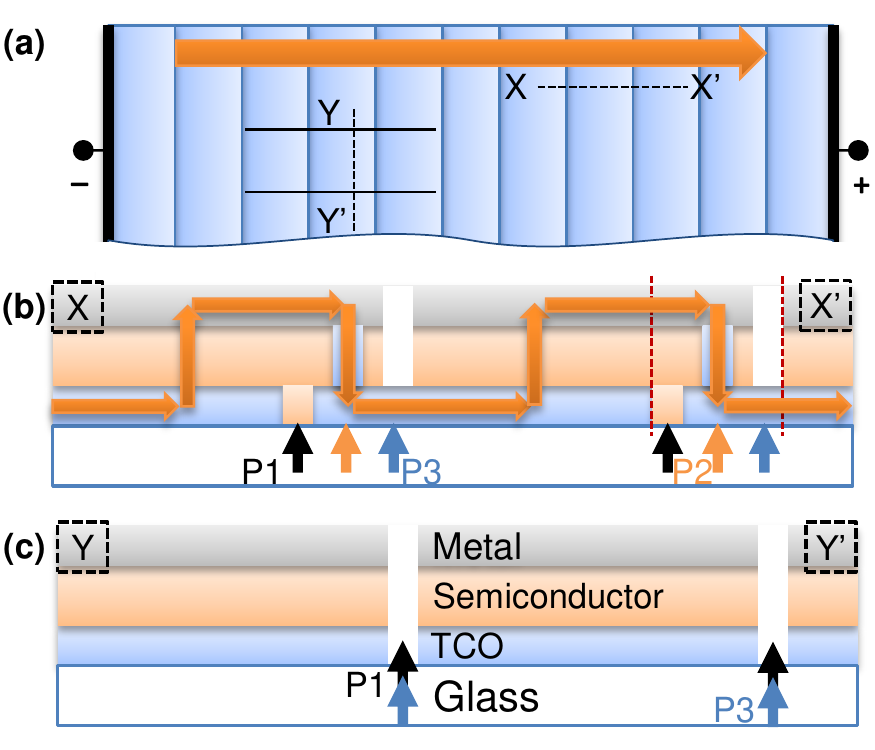}
\caption{(a) Schematic showing the top view of series connected cells in a module with direction of current flow as shown. (b) Side view of the TFPV module along XX’, showing the series connection created using adjacent P1/P2/P3 laser scribes (arrows). (c) Cross section along YY’ showing two isolating scribes created by overlapping the P1 and P3 lasers, to cut-off the region between the scribe lines.}
\label{fig4}
\end{figure}

\subsection{Scribing Isolation Methods}

From the simulations in section III.A, we had seen that the significant reduction in module efficiency due to shunting arises from the non-local effect of shunted sub-cells on their neighbors. Therefore, a significant portion of shunt induced losses can be avoided if this interaction between the neighboring sub-cells can be suppressed. Fortunately, the laser scribing method used for creating the series connections in TFPV modules (see cross section XX’ in Fig. \ref{fig4}(b)) can be used to create insulating (open circuit) scribes, which will isolate the shunted region (see cross section YY’ in Fig. \ref{fig4}(c)). Note that while the series connections (Fig. \ref{fig4}(b)) are created using successive P1-P2-P3 laser scribes adjacent to each other; the isolating scribes are created by superposing P1 and P3 lasers to disconnect all layers (Fig. \ref{fig4}(c)), see \cite{Haas} for details.

\subsubsection{Periodic Lengthwise Scribing}

A straightforward method for shunt isolation using scribing involves creating full length isolating scribes in the direction of current flow in a TF module. These vertical scribes divide the rectangular cells in the module into columns of smaller width, each of which are connected to the common bus bar (see the schematic in Fig. \ref{fig5}(a)). This ensures that the effects of shunted sub-cells are restricted to fewer neighboring sub-cells (one for the case shown in Fig. 2(a)). This approach does not require identification of the shunt locations beforehand. The scribe lines, however, need to span the entire module length, which ensures that regardless of the shunt location, the number of neighboring sub-cells in parallel it can influence is limited. Note that this full length scribing isolation at the sub-module level has been demonstrated earlier using mechanical scribing for edge shunt isolation \cite{Pern2011}. In this work, we propose to extend this approach, and create periodic scribes for constraining shunt effect at all locations.

\subsubsection{Selective Scribing Isolation}

It is apparent from the previous section that lengthwise scribing will require very long scribe lines for a typical sized TFPV module. It would therefore be beneficial to first identify the shunts, and scribe only around the shunted sub-cells. There are a variety of imaging methods, which can be used to identify the spatial location of shunts. These can either be contactless methods like PL \cite{Johnston2010a} or EBIC \cite{Yacobi1984} imaging, contact methods like EL \cite{Helbig2010} or DLIT \cite{Rosch2012} imaging, or a combination thereof. Most of these techniques have been studied extensively, and are being actively adapted for inline metrology purposes \cite{Hoppe2010}. Once the images have been used to identify shunt location, an image processing routine can automatically extract their coordinates \cite{Miranda2010}, which can then be used to position the scribe heads in the right location.

\begin{figure}[t]
\centering
\includegraphics[width=3.49in]{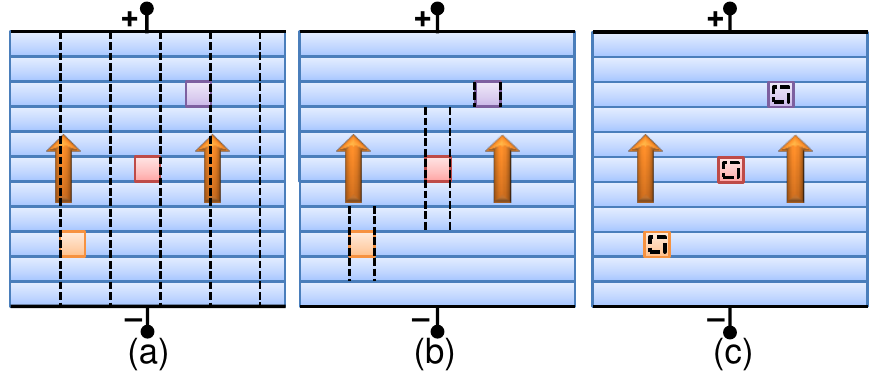}
\caption{Sub-module schematics showing the three shunted sub-cells, compare the different scribing isolation schemes (isolating scribes shown with dashed lines). (a) Full length isolating scribes at periodic intervals, in the direction of current flow (arrows). (b) Partial isolation using parallel scribes, after shunt location detection, with different scribe lengths providing different degree of isolation. (c) Full isolation of shunted sub-cells using box scribes encircling the shunted region inside the sub-cell.}
\label{fig5}
\end{figure}

Once the shunt locations are known, we can either choose to partially isolate, or fully isolate the shunted sub-cell. The partial isolation is created using 2 isolating scribe lines in parallel to the direction of current flow, either side of the shunted sub-cell (schematic in Fig. \ref{fig5}(b)). These lines ensure that the shunted sub-cell is disconnected from its neighboring sub-cells in parallel, while maintaining the current continuity through the series connected cells. Note that the sub-cells in series with the shunted ones, which are inside the isolating scribes, will now operate away from their MPP voltage, due to the condition of current continuity. This partial isolation, however, does not require very precise determination of shunt location, or positioning of the scribing heads, and will be easier to implement in practice. 

The second option isolates the shunt fully by enclosing the shunt using isolating scribes (see schematic in Fig. \ref{fig5}(c)). This ensures full isolation, at the cost of greater area penalty, as all the area enclosed by the scribe lines is disconnected from the module and cannot produce output power. This requires a more precise identification of shunt locations, as well as more accurate positioning of scribe heads, in order to ensure that the shunt is in fact enclosed by the scribes. The shunts in this case, however, are fully isolated, and do not affect any sub-cells in series or parallel after scribing.

From the previous discussion, it appears that this scribing isolation approach simply exchanges shunt losses with dead area loss. In section III.A, however, we saw that the largest shunts have a disproportionate impact on module efficiency. And, due to the log-normal nature of shunt statistics, only a small fraction of sub-cells are heavily shunted \cite{Dongaonkar2013b}. Therefore, we only need to isolate a small fraction of the total shunts (largest ~3\%),  in order to achieve significant performance recovery. This is the reason why the scribing isolation can recover significant shunt losses, at a relatively small increase in dead area loss due to few extra scribe lines.

\section{Comparison of Scribing Isolation Schemes}

\begin{figure}[t]
\centering
\includegraphics[width=3.3in]{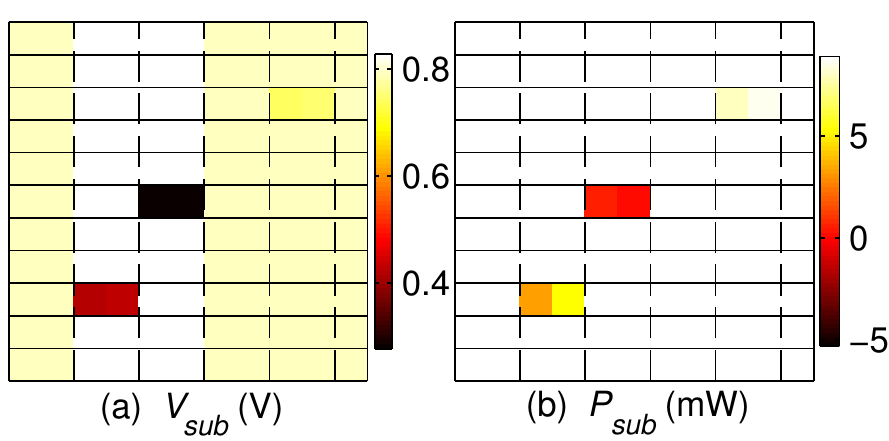}
\caption{(a) Color plot of sub-module $ V_{sub} $ (color bar in $ V $), in presence of full length isolating vertical scribe lines (dotted), showing that the operating voltages of the neighboring sub-cells in parallel improve with scribing. The non-shunted sub-cells inside the scribes are slightly forward biased to ensure series connected conduction. (b) Distribution of $ P_{sub} $ values (color bar in $ mW $), showing that majority of sub-cells now operate at MPP, and the power output in shunted sub-cells is also improved.}
\label{fig6}
\end{figure}

\subsection{Sub-module Operation with Shunt Isolation}

In this section we revisit our illustrative example of an $ 11\times11cm^2 $ sub-module with 3 shunts. We will apply the various scribing schemes on this sub-module, and analyze the resulting voltage and power distributions carefully to gain insight into how these schemes improve module performance.

\subsubsection{Periodic Lengthwise Scribing}

We first analyze the case of full length periodic scribe lines at the sub-module level, as shown in Fig. \ref{fig5}(a). The $ V_{sub} $ and $ P_{sub} $ distributions obtained from the simulations are shown in Fig. \ref{fig6}(a) and Fig. \ref{fig6}(b), respectively. Note that in this scenario, the effective range of all shunted sub-cells is restricted to only one neighboring sub-cell in parallel. Consequently, the operating voltages and power outputs of the `good' sub-cells improves in general. Note, however, that due to series connection, the voltage of each column of sub-cells (between two isolating scribes) must be equal. Therefore, the operating points of the sub-cells in series to the shunted sub-cells now must move to slightly higher $ V_{sub} $ values (see columns containing shunted sub-cells marked (2) and (3)). In spite of this limitation, note that the blocking of parallel conduction paths, results in all sub-cells producing power (i.e. $ P_{sub}>0 $ for all sub-cells in Fig. \ref{fig5}(b)). Therefore, we can see that this approach will lead to improved sub-module output.

\subsubsection{Selective Scribing - Partial Isolation}

\begin{figure}[t]
\centering
\includegraphics[width=3.3in]{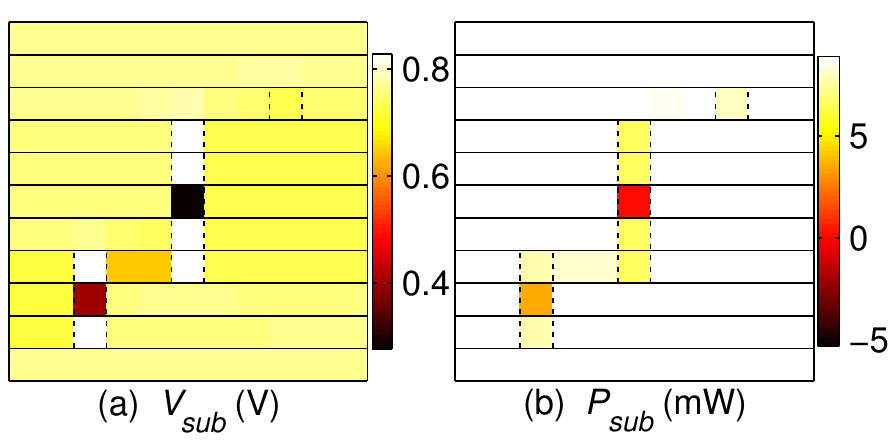}
\caption{(a) Color plot of sub-module $ V_{sub} $ (color bar in $ V $), in presence of isolating vertical scribe lines of different length (dotted), showing that the operating voltages of the neighboring sub-cells in parallel are back to MPP values. The non-shunted sub-cells inside the scribes are slightly forward biased to ensure series connected conduction. (b) Distribution of $ P_{sub} $ values (color bar in $ mW $), showing that majority of sub-cells now operate at MPP, and the power output in shunted sub-cells is also improved. Both these plots show that longer scribes allow better isolation, and better sub-cell characteristics.}
\label{fig7}
\end{figure}

If the shunt positions are known beforehand, we can use parallel isolating scribes on either side, of varying lengths ($ L_{scribe} $). This ensures that the non-shunted sub-cells are not unnecessarily scribed. Fig. \ref{fig7}(a) shows the color plot $ V_{sub} $ values from the sub-module simulation, in presence of the scribes shown in Fig. \ref{fig5}(b). Note that the operating points of sub-cells in parallel to shunted sub-cells improve close to MPP values. The $ V_{sub} $ values of the sub-cells in series to the shunted sub-cells however, now increase slightly to ensure total sub-module voltage being same. The sub-cell power outputs shown in Fig. \ref{fig7}(b) also show the reduced impact of shunted regions on neighboring sub-cells in parallel, at the cost of slight reduction in $ P_{sub} $ values for the sub-cells in series with the shunted sub-cells. Finally, note that the length of scribe lines used for isolation affects the degree of isolation, and accordingly, the efficiency improvement obtained, at the cost of higher dead area loss.

\subsubsection{Selective Scribing - Full Isolation}

\begin{figure}[t]
\centering
\includegraphics[width=3.3in]{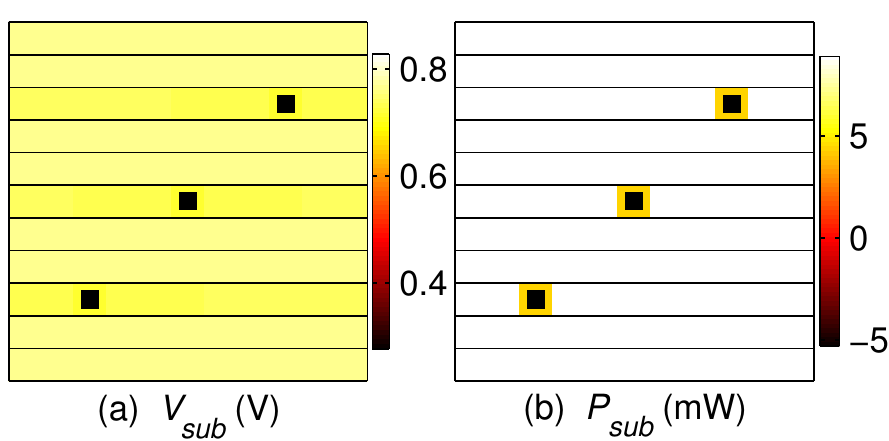}
\caption{(a) Color plot of sub-module $ V_{sub} $ (color bar in $ V $), with full isolation of the three shunts (black squares), showing that the operating voltages of almost all sub-cells are restored to MPP values. (b) Distribution of $ P_{sub} $ values (color bar in $ mW $), showing that all non-shunted sub-cells now operate at MPP, and the power output in shunted sub-cells is also improved considerably. These plots show that the area mismatch caused by full isolation has a limited impact on sub-module performance as a whole.}
\label{fig8}
\end{figure}

Finally, we analyze the case with full shunt isolation, achieved by enclosing them completely using isolating scribe (as shown in Fig. \ref{fig5}(c)). This approach  leads to  slightly higher dead area loss, since the region inside the scribe lines will not contribute to the output. The sub-module simulation results with full isolation are shown in Fig. \ref{fig8}. Note that the isolation around the shunts reduces the voltage discrepancy across the different sub-cells considerably (Fig. \ref{fig8}(a)). The isolated region is shown by dark squares, as it is completely cut-off from the module output. Correspondingly, all the non-shunted sub-cells now operate near their MPP (Fig. \ref{fig7}(b)). Depending on the total area inside the scribed region ($ A_{scribe} $), some portion of the shunted sub-cells will still be generating power (see Fig. \ref{fig7}(b)), and the fraction of area lost to scribing per cell ($ A_{scribe}/A_{sub} $) will determine the eventual benefit of this scheme.

\subsection{Scribing Isolation for Module Efficiency Improvement}

From the previous discussion, we can see that various scribing techniques improve the module performance, by isolating the shunts to varying degrees, but the trade-off in terms of dead area loss is also different. We now analyze these tradeoffs more quantitatively, for typical size commercial modules with parasitic shunt distribution. We consider the highest efficiency a-Si:H module ($ \eta=7.22\% $) from the CDF in Fig. \ref{fig3}, which has a certain distribution of shunts on the surface responsible for an efficiency loss of $ \sim2\% $ (absolute). We will apply the various scribing schemes to this particular module to evaluate the improvement in efficiency obtained. This will allow us to evaluate the effectiveness of these schemes, under realistic circumstances. In addition to the typical module and sub-cell parameters used in section III.B, we use a realistic value for isolating scribe lines to be $ 150\mu m $ \cite{Haas}. Note that the dead area loss in the scribed sub-cells is accounted for in the SPICE simulation by modifying the active area of each cell, based on the presence of isolating scribes. This also ensures that any effects due to mismatches in sub-cell IV are accounted for when evaluating the effect of scribing schemes.

\begin{figure}[t]
\centering
\includegraphics[width=3.49in]{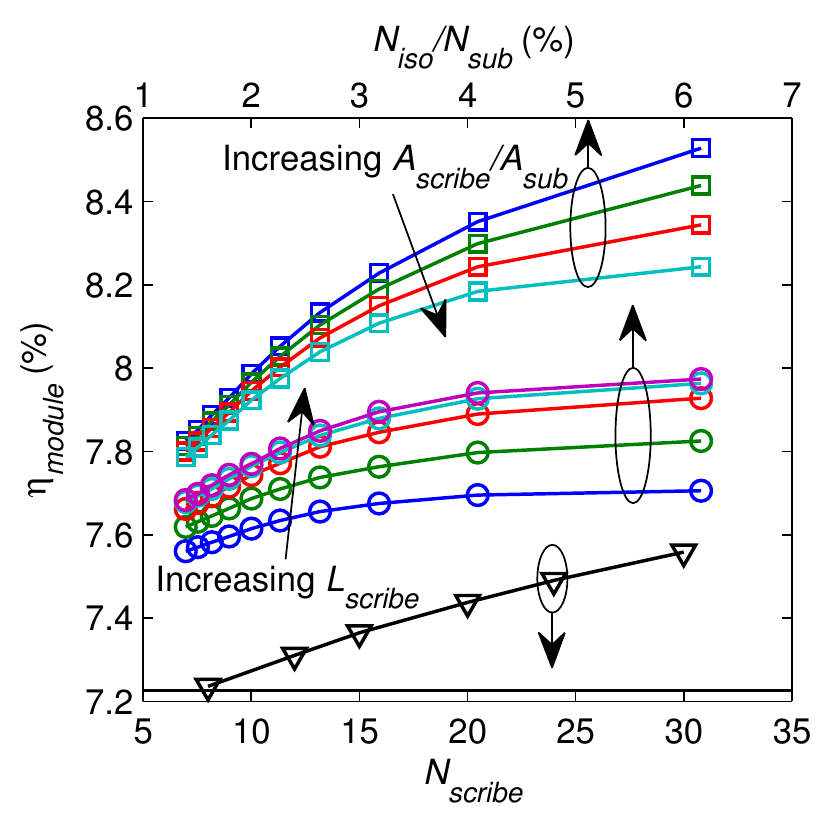}
\caption{Plot comparing the efficiency of a typical 7.22\% a-Si:H module with parasitic shunts (horizontal line), after the module is treated using different scribing isolation techniques. For the lengthwise isolating scribes (triangles), without detection, the efficiency improves with more vertical scribe lines ($ L_{scribe} $), but the efficiency gains remain modest due to large dead area penalty. The efficiency enhancements are significantly better if selective scribing is used to isolated more and more shunted sub-cells (increasing $ N_{iso}/N_{sub} $) for both partial (circles) and full (squares) isolation case. In case of partial isolation, the efficiency improves for larger $ L_{scribe} $, due to better isolation, but ultimately limited by dead area losses (squares). For full isolation, the dead area losses are determined by the ratio $ A_{scribe}/A_{sub} $, which limits the efficiency enhancements obtained due to reduced shunt effects.}
\label{fig9}
\end{figure}

For the lengthwise scribing, it is apparent from Fig. \ref{fig6} that an increase in the number of vertical scribes will improve the module efficiency by containing the effect of shunted sub-cells. This, however, requires very long scribe lines, so the dead area penalty also increases, and the incremental gains in module efficiency with increasing number of scribes ($ N_{scribe} $) is positive but modest. As shown by the triangles in in Fig. \ref{fig9}, the module efficiency improves to $ \sim7.6\% $ for 30 full length vertical scribes. This efficiency improvement, however,   rises slowly with increasing $ N_{scribe} $ to larger values.

In case of targeted shunt isolation after determination of shunt locations, we need to choose the number of shunted sub-cells to isolate. This is also determined by the accuracy of the detection scheme. As expected as we isolate more and more shunted sub cells, out of the total $ N_{sub} $ (i.e. increasing $ N_{iso}/N_{sub} $ in Fig. \ref{fig9}), the module efficiency improves significantly (see circles for partial isolation, and squares for full isolation in Fig. \ref{fig9}). As expected, the improvements are limited by increased dead area loss with additional scribing, and the curves saturate for larger $ N_{iso}/N_{sub} $ values. Interestingly, note that using the prior detection of shunts, it is possible to get $ \eta_{module} \geq 7.6\% $ by partially or fully isolating just the largest 1\% shunt defects. This is a consequence of the heavy tailed log-normal shunt distribution, which means that few largest shunts dominate the overall efficiency loss at the module level (see \cite{Dongaonkar2013b} for details). Therefore, we can obtain significant improvements by focusing on these most critical parasitic defects.

In case of partial isolation technique another optimization parameter is the length of isolating scribe $ L_{scribe} $, used for each shunted sub-cell. As expected from Fig. \ref{fig7}, longer scribe lines provide better isolation, and the efficiency improves for the same $ N_{iso}/N_{sub} $ value by increasing $ L_{scribe} $ from 1 to 9 cm in steps of 2 cm (see circles in Fig. \ref{fig9}). Understandably, the dead area loss limits these gains, and therefore, the module efficiency for $ L_{scribe}=7cm $ vs. $ L_{scribe}=9cm $ are virtually identical. 

In case of full isolation of shunts, the efficiency gains will be limited by the ratio between scribed sub-cell area, and total sub-cell area $ A_{scribe}/A_{sub} $. In a practical scenario, this ratio will depend on the accuracy of the shunt detection, and scribing head positioning. We, therefore, vary the dead area loss ($ A_{scribe}/A_{sub} $) from 20\% to 80\% (see squares in Fig. \ref{fig9}), and show that regardless of the dead area loss associated with full scribing, the full shunt isolation provides highest module efficiency out of the 3 suggested techniques. This is because, by fully isolating the shunts this approach removes the effect of shunted sub-cells on all its neighbors (series or parallel). Furthermore, the random distribution of shunts on a module surface ensures that the mismatches introduced by scribing area loss are not significant.

\subsection{Scribing Isolation and Module Yield}

Based on the discussion so far, we observe that all three scribing techniques enable improvement in module efficiency. We now show that these results hold true in a statistical sense, and compare the change in module efficiency distribution for different scribing methods. In order to compare the techniques statistically, we repeat the Monte-Carlo simulation of module efficiency in section III.B, applying the different scribing techniques for each module. For the lengthwise periodic scribing, we choose full length scribes every 5 cm; and, for the selective scribing schemes we choose$ N_{iso}/N_{sub}=3\% $. For the partial isolation method, $ L_{scribe}=5cm $, and $ A_{scribe}/A_{sub}=0.4 $ for the full isolation case. These numbers are chosen because, from the analysis in the previous section, we note that the incremental gain in module efficiency with scribing more shunts or using longer scribe is minimal. Moreover, these values are reasonable from a practical standpoint of implementing this strategy in a production environment. In practice, these quantities need to be optimized, depending upon the technology of choice, the shunting and scribing parameters of the process, and the required efficiency improvement.

\begin{figure}[t]
\centering
\includegraphics[width=3.49in]{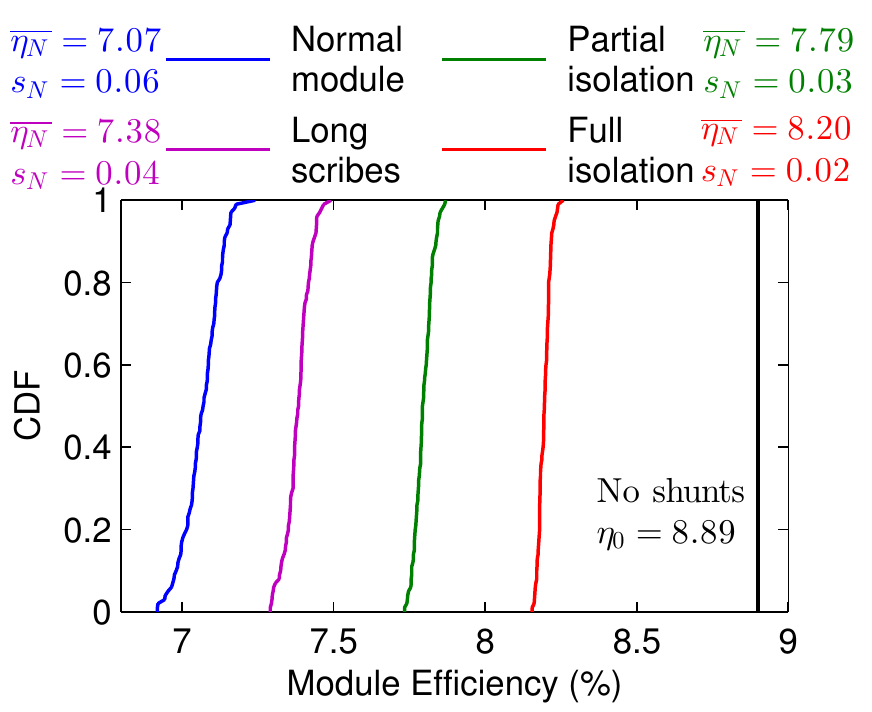}
\caption{CDF plots of module efficiency distribution for typical case (blue), lengthwise periodic scribing (magenta), partial shunt isolation after detection (green), and full shunt isolation after detection (red). The improvement in mean efficiency ($ \bar{\eta}_N $ ) ̅, and reduction in standard deviation $ s_N $ are apparent from the values shown in inset. The full isolation of shunts can yield efficiency values close to the maximum possible module efficiency (black).}
\label{fig10}
\end{figure}

Fig. \ref{fig10} compares the CDFs of module efficiency without any scribing vs. each scribing method. Note that as expected from previous sections, the mean module efficiency $ \bar{\eta} $ improves by $ \sim0.3\% $ (absolute) for lengthwise periodic scribing, $ \sim0.7\% $ (absolute) for partial scribing and $ \sim1.1\% $ (absolute) for full scribing isolation method. Interestingly, note that simply by isolating only 3\% of the shunts, and incurring significant area penalty for each isolating box, we can bring the module efficiency very close to ``ideal'' module efficiency of 8.9\%, which is obtained for the case with no shunt defects. Equally important, the scribing isolation of shunts also leads to reduced standard deviation of the module efficiency distribution, see Fig. \ref{fig9}. This means that in addition to the overall module efficiency improvement, the scribing techniques also reduce the performance variability associated with random shunt formation. This in turn will improve the overall yield of the manufacturing process, which is an important factor in PV manufacturing cost.

\section{Conclusion}

In this paper we demonstrate an in-line method for module efficiency improvement in monolithic TFPV modules, by isolation of parasitic shunt defects using laser scribing. The approach utilizes state of the art techniques of shunt identification, and laser scribing, to isolate a small fraction of the largest shunts in the module. We describe two techniques, which involve either partial isolation, or full isolation of the shunt defects. We used 2D, self-consistent, SPICE circuit simulations of the modules in presence of shunts, to compare these techniques quantitatively and illustrate the involved optimizations. We demonstrate using typical parameters that these scribing techniques can recover more than half of the module efficiency loss caused by parasitic shunts. 

Finally, we would like to emphasize that these techniques are equally applicable to all monolithic TFPV technologies, and owing to the similarities in shunt statistics \cite{Dongaonkar2013b}, are expected to yield similar performance gains at the module level. Indeed, while this analysis is presented for effect of shunt leakage, the isolation scheme will be equally beneficial for any other type of parasitic localized leakage/non-uniformity, e.g., weak diodes \cite{Koishiyev2009e}, which can affect their neighboring regions adversely. The implementation of these techniques in realistic production line, however, will have to be analyzed on a case by case basis; including the degree of shunt or other leakage losses, and economics of implementing an extra process step. In this regard, the scribing techniques is advantageous, as they only require the state-of-the-art manufacturing tools for implementation, and only add one extra, post-process step to the flow. 

We believe that the potential advantages of these schemes for module efficiency and process yield enhancement, coupled with relatively straightforward implementation, and applicability to different thin film technologies \cite{Dongaonkar2013a}, will motivate the PV community to explore these approaches in production environments.

\section*{Acknowledgment}

We would like to acknowledge the Network for Computational Nanotechnology for the computational resources used in this work. We would like to thank B. J. Stanbery for helpful discussions.

\end{document}